\documentclass{PoS}
\PoS{PoS(LAT2005)313}
\title{Real-time dynamics of proton decay }
\ShortTitle{Real-time dynamics of proton decay }
\author{\speaker{Dmitri Grigoriev}
\\     Dept. of Mathematical Physics, NUI Maynooth, Co. Kildare, Ireland\\
        E-mail: \email{dima@thphys.may.ie}}

\abstract{
Substituting Skyrmion for nucleon, one can potentially see --- in real
time --- how the monopole is catalysing the proton (or neutron) decay, and
even obtain a plausible estimate for catalysis cross-section. Here we
discuss the key aspects of a practical implementation of such approach
and demonstrate how one can overcome the main technical problems:
Gauss constraint violation and reflections at the boundaries.}

\FullConference{XXIIIrd International Symposium on Lattice Field Theory\\
                 25-30 July 2005\\ Trinity College, Dublin, Ireland}

\begin{document}
\newcommand{\gsim}{\raisebox{-0.8ex}{\mbox{$\stackrel{\textstyle>}{\sim}$}}}
\newcommand{\lsim}{\raisebox{-0.8ex}{\mbox{$\stackrel{\textstyle<}{\sim}$}}}
\newcommand{\re}[1]{(\ref{#1})}
\newcommand{\dl}{^{\rm dl}}
\newcommand{\dll}{_{\rm dl}}
\newcommand{\msp}{M_{\rm sph}}
\newcommand{\esp}{E_{\rm sph}}
\newcommand{\rsp}{R_{\rm sph}}
\newcommand{\dsp}{D_{\rm sph}}
\newcommand{\msk}{M_{\rm sk}}
\newcommand{\rsk}{R_{\rm sk}}
\newcommand{\esk}{E_{\rm sk}}
\newcommand{\dsk}{D_{\rm sk}}
\renewcommand{\max}{M_{\rm axial}}
\newcommand{\mm}{M_{\rm mon}}
\newcommand{\ma}{m_{\rm A}}
\newcommand{\mps}{m_\Psi}
\newcommand{\dm}{D_{\rm mon}}
\newcommand{\mv}{m_{\rm V}}
\newcommand{\mh}{m_{\rm H}}
\newcommand{\agut}{\alpha_{\rm GUT}}
\newcommand{\eins}{1\hspace{-0.56ex}{\rm I}}

\newcommand{\1}{\cite{brtz}}
\newcommand{\fax}{f_\infty}
\newcommand{\be}{\begin{equation}}
\newcommand{\ee}{\end{equation}}
\newcommand{\bea}{\begin{eqnarray}}
\newcommand{\eea}{\end{eqnarray}}
\newcommand{\const}{\mathop{\rm const}\nolimits}

\section{Introduction}

The monopole catalysis of proton decay \cite{R,C} is a complicated process
involving several different physical scales. Although the physics of
how fermions interact with the monopole is rather well understood, the
net proton decay cross-section remains unknown, mainly because the proton decay
actually happens via a combination of quite different physical mechanisms. At
first, spatially separated monopole and proton interact at relatively
large distances; then, the monopole starts to interact with individual
quark states inside the proton, whose internal structure is defined by
non-perturbative QCD phenomena.

An interesting possibility to study both kinds of processes is
provided \cite{CW} by the Skyrme theory \cite{skyrme}. However, even
though the monopole catalysis of the Skyrmion decay is essentially a
classical process, its cross-section is also not known. Recently it
has been suggested \cite{brtz,Brihaye:2002nz} to study this process in
a system of the Skyrmion interacting with the 't Hooft-Polyakov
monopole \cite{mon-t,mon-p}. Since the latter is a spatially extended
object, this makes the model free from singularities and greatly
simplifies its numerical study. In the papers  \cite{brtz,Brihaye:2002nz}, the
spherically-symmetric monopole-Skyrmion geometry has been
comprehensively studied in quasistatic limit. In particular, the decay
path of the Skyrmion has been identified, and it has been shown that
nothing prevents the Skyrmion from promptly losing almost all of its
energy in the monopole background.

The natural next step is to study the classical dynamics of the model
suggested in  \cite{brtz,Brihaye:2002nz}. In this talk we discuss the main technical aspects of
such a study, while the physical results will be published
elsewhere \cite{tobe}. In particular, we concentrate on a problem of gauge
invariance violation in real-time classical simulations.

\section{The model}

The model\footnote{For a detailed discussion of the structure and the
properties of the model see the Refs.~\cite{brtz,Brihaye:2002nz}.} consists of the Skyrme
field $U$ coupled to two gauge fields $A_{\mu}$ (group $SU(2)_L$) and
$B_{\mu}$ (group $SU(2)_R$). The Higgs sector has two Georgi-Glashow
Higgs fields $\Phi_A$ and $\Phi_B$ which break the $SU(2)_L$ and
$SU(2)_R$ symmetries down to $U(1)_L$ and $U(1)_R$, respectively, and
a third Higgs field $\Psi$ which breaks the axial subgroup of the
remaining $U(1)_L \times U(1)_R$. These requirements fix the scalar
field group representations completely: $ U \in SU(2)$ ,
$U,\Psi:\;\;({\bf 2}, {\bf \bar{2}})$, $\Phi_A:\;\; ({\bf 3}, {\bf
1})$, $\Phi_B : \;\; ({\bf 1}, {\bf 3})$.

The action for the model is \begin{eqnarray}
     S &=& \int~d^4x~\left[ -\frac{1}{2g^2} \mbox{Tr}(F_{\mu\nu}^2)
 -\frac{1}{2g^2} \mbox{Tr}(G_{\mu\nu}^2)\right] \nonumber \\
  &-& \int~d^4x~\left[
\frac{1}{2} \mbox{Tr}(D_{\mu}\Phi_A)^2 + 
\frac{1}{2} \mbox{Tr}(D_{\mu}\Phi_B)^2
\right]
\nonumber \\
&-&  \int~d^4x~\left[
\frac{\lambda}{4} 
\left(\frac{1}{2} \mbox{Tr}(\Phi_A^2) + v^2\right)^2
+  \frac{{\lambda}}{4} 
\left(\frac{1}{2}\mbox{Tr}(\Phi_B^2) + v^2\right)^2
\right] \nonumber \\
&+& \int~d^4x~\left[\frac{1}{2}\mbox{Tr}(D_{\mu}\Psi^\dagger
D_{\mu}\Psi) 
+
\frac{\tau}{4}\left(\frac{1}{2}\mbox{Tr}(\Psi^\dagger\Psi)-\zeta^2\right)^2
\right]\nonumber\\
&+& \int~d^4x~\left[  
- \frac{F_{\pi}^2}{16} \mbox{Tr}(U^{\dagger} D_{\mu} U)^2
 + 
\frac{1}{32e^2}\mbox{Tr}([U^{\dagger}D_{\mu}U,U^{\dagger}D_{\nu}U]^2)
\right] \nonumber \\
&+& \Gamma_{WZW}
\label{action}
\end{eqnarray}
Here $F_{\mu\nu}$ and $G_{\mu\nu}$ are the field strengths
of $A_{\mu}$ and $B_{\mu}$, the fields $\Phi_A$ and $\Phi_B$
are $2 \times 2$ matrices from the algebras of $SU(2)_L$ and 
$SU(2)_R$, respectively,
\begin{eqnarray}
    D_{\mu}U &=& \partial_{\mu} U + A_{\mu}U - UB_{\mu}
      \nonumber \\
    D_{\mu} \Phi_A &=& 
     \partial_{\mu}\Phi_A + [A_{\mu}, \Phi_A]
\nonumber \\
D_{\mu} \Phi_B &=& 
     \partial_{\mu}\Phi_B + [B_{\mu}, \Phi_B]
\end{eqnarray}
and $F_{\pi}$ and $e$ are the pion decay constant and the Skyrme
constant. The new Higgs field $\Psi$
is a $2 \times 2$ complex matrix with covariant derivative
identical to that  of the Skyrme field, $$D_{\mu}\Psi = \partial_{\mu}
\Psi + A_{\mu}\Psi - \Psi B_{\mu}$$ Finally, the 
Wess--Zumino--Witten term
$\Gamma_{WZW}$ can be safely ignored \cite{brtz} in what follows.

The most general spherically
symmetric Ansatz consistent with the symmetries \cite{brtz,Brihaye:2002nz} of the action \re{action} is
\begin{eqnarray}
A_0 &=& -B_0=-{i\over 2}\left( \frac{a_0 (r)}{r} \right)\hat x \cdot \vec
\sigma \nonumber \\
A_i &=& -{i\over 2}
\left[ \left( \frac{a_1 (r)-1}{r}\right)\varepsilon_{ijk}\sigma_j \hat x_k
+\left( \frac{a_2 (r)}{r} \right) 
(\sigma_i -\hat x_i \hat x \cdot \vec \sigma )
+\left( \frac{a_3 (r)}{r} \right) 
\hat x_i \hat x \cdot \vec \sigma \right] \nonumber \\
B_i &=& -{i\over 2}
\left[ \left( \frac{a_1 (r)-1}{r}\right)\varepsilon_{ijk}\sigma_j \hat x_k
-\left( \frac{a_2 (r)}{r} \right) 
(\sigma_i -\hat x_i \hat x \cdot \vec \sigma )
-\left( \frac{a_3 (r)}{r} \right) 
\hat x_i \hat x \cdot \vec \sigma \right]\:
\nonumber \\
\Phi_A &=& \Phi_B = i v h(r) \hat x \cdot \vec \sigma
\nonumber \\
U&=&\eins\cos f(r) +i\hat x \cdot \vec \sigma \sin f(r)\;\;\; ,\;\;\;
\Psi=\zeta[ih_1(r)\hat x.\sigma+h_2(r)\eins]
\label{6m**}
\end{eqnarray}
where $\hat x$ is the unit radius-vector.

Substituting the ansatz \re{6m**} into \re{action} and using the  $A_0 = B_0 = 0$ gauge to get the proper Hamilton real-time evolution, one obtains
spherically-symmetric action which remains invariant to the following
static gauge transformations which are the remains of the axial $U(1)$:
\begin{eqnarray}
h_2+ih_1 &\to& (h_2+ih_1) e^{i\alpha(r)} \nonumber \\
a_1+ia_2 &\to& (a_1+ia_2) e^{i\alpha(r)} \nonumber \\
    a_3 &\to& a_3 + r \partial_r \alpha(r) \nonumber \\
    f &\to& f + \alpha(r)
\label{restgauge}
\end{eqnarray}
(note that $h_2+ih_1$ rotates identically to $e^{if}$, once the 
fields $\Psi$
and $U$  are in the same group representation). This remaining gauge invariance becomes a major obstacle in real-time simulations (see below).

The classical equations of motion can be obtained from the
spherically-symmetric action by the standard variational procedure. Of
particular importance for what follows is the Gauss constraint
corresponding to variation over $a_0$:
\be
\Delta=-\frac1{2g^2}\left[\partial_r\dot a_3+\frac{\dot a_3}{r}\right]
+\frac{a_1\dot a_2-a_2\dot a_1}{g^2r} 
+\left[\frac{F_{\pi}^2}{8}r^2+\frac1{8e^2}(a_1\sin f-a_2\cos f)^2\right]\frac{\dot f}{r}
-\zeta^2r(h_1\dot h_2-h_2\dot h_1)=0
\label{ta0}
\ee
The constraint \re{ta0} is preserved by the equations of motion as long as their gauge invariance is not violated.

\section{Numerical study}

In the $A_0 = B_0 = 0$ gauge the classical equations of motion
obtainable from the action \re{action} represent a system of
2$^{\rm nd}$-order hyperbolic partial differential equations in (1+1) dimensions which
can be solved numerically by e.g. a staggered
leapfrog scheme. Choosing different initial configurations, one can
study several physically interesting cases of the Skyrmion decay. For
example, starting from a normal vacuum Skyrme solution (so-called
``bare'' Skyrmion, see \cite{brtz,Brihaye:2002nz}), one can simulate
the decay of a physical nucleon. By choosing the initial profile for
$f(r)$ corresponding to the ``thin'' Skyrmion (a static
solution \cite{brtz,Brihaye:2002nz} with unit baryon charge and
minimal energy in the monopole background), it becomes possible to
check \cite{tobe} how close is its actual decay path to  the quasistatic paths
found in the Refs.\cite{brtz,Brihaye:2002nz}. However, any such study is
facing two serious difficulties related to the lack of gauge-invariant
lattice action for the model and the need for fully absorbing boundary
conditions.

\subsection{Lattice action}

To preserve the physical structure of continuum theory, its symmetries
must be retained in the discretised lattice version, and, in particular,
the symmetries to local gauge transformations. Unfortunately, we are
not aware of any gauge-invariant lattice action for gauged Skyrme
models, including the model used in the present study. The use of
gauge-noninvariant lattice action, generally speaking, results in
domination of nonphysical states in generated field
configurations. However, in real-time classical evolution the mixing
of physical and nonphysical states can be rather well controlled.

Since the only reason for gauge invariance violation are the
discretisation errors, for reasonably small lattice spacing the
magnitude of the violation remains rather small. Assuming that the
evolution starts from a physical state which obeys the Gauss
constraint \re{ta0}, the system initially stays close to the physical
manifold and can be projected back onto it after certain amount of
time. Physically, the projection procedure involves the minimisation of the quantity 
\be Q=\int dx\;\Delta^2 \ee
where $\Delta$ is the constraint value (the left-hand side of
\re{ta0}). Generally speaking, any minimisation technique can be used
here, with no specific approach favoured by physical reasons as long
as the accumulated constraint violation remains small. In the present study we
use a variation \cite{Grigoriev:1989ub} of the steepest descent
method, also called sometimes as the Langevin cooling.

It is worth noting that in real-time classical simulations the projection procedure is inevitable
\cite{Grigoriev:1989ub} even with a
perfectly gauge-invariant lattice action. The reason is that the gauge
invariance will inevitably be broken by computer roundoff errors. Even
though the latter remain tiny ($O(10^{-15})$ for the standard double
precision), the constraint violation accumulates in the course of
evolution, which results in exponential divergency of constraints. To
remain close to the physical manifold for indefinite period of time,
one still has to periodically project the system onto it, although for
gauge-invariant lattice schemes the constraint violation can be
maintained at dramatically lower level.

Failure to keep constraint violation under control inevitably leads to an
explosion of the numerical scheme, whether a gauge-invariant one or not. For
the model at hand this is demonstrated at the Figure~1, where the
system energy is plotted against the time along with the magnitude of
the volume-averaged constraint value $<\Delta>_{\rm vol}$.

\begin{figure} 
\centerline{\includegraphics[width=.91\textwidth]{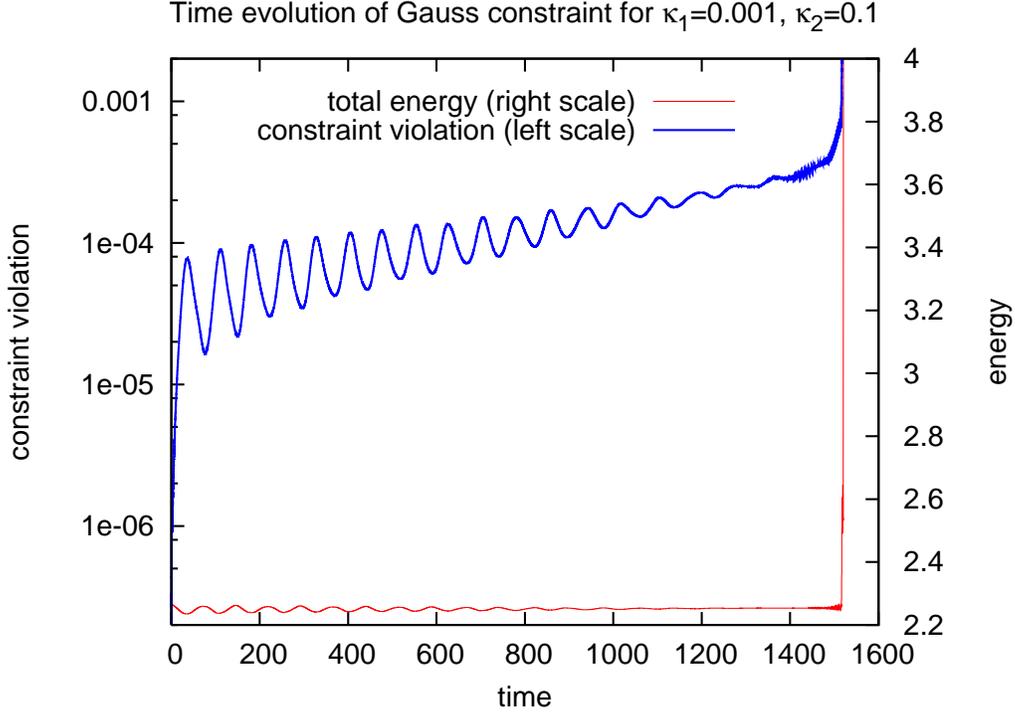}}
\caption{ Constraint violation inevitably breaks stability of the
numerical evolution, with the Gauss constraint and total energy
exploding simultaneously after the former exceeds a certain critical
level.  Of course, the numerical scheme also violates the energy
conservation (as demonstrated by the small oscillations of energy),
but in non-gauge models \cite{gr} this doesn't break the stability.
Notations: $ \kappa_1 = F_{\pi}^2/8 v^2$, $\kappa_2 =g^2/64e^2$; for dimensionless units used see
Ref.\cite{Brihaye:2002nz}.  }
\label{fig1} 
\end{figure}

\subsection{Absorption at the boundaries}

In physical case, the energy of the decaying proton is released to the
spatial infinity. In numerical studies, this energy should be either
absorbed at the boundaries or  allowed to pass through the
boundaries. For massless waves this can be easily implemented by the
standard outgoing-wave boundary conditions. However, in more realistic cases,
including the present study, it turns out to be very difficult to
achieve full absorption at the boundaries, simply because any
modification of the theory near the boundaries introduces spatial
inhomogeneities which scatter the waves back to centre.
Fortunately, in (1+1)-dimensional case it is still possible to
overcome the problem by brute force, taking a sufficiently large
physical volume, so that the reflected wave wouldn't come back to the
origin before the end of the run, but this obviously cannot be done for higher spatial dimensions.

\section{Conclusions}

Despite the problems discussed above, the monopole catalysis of
Skyrmion decay can be successively studied \cite{tobe} in the
spherically-symmetric case where one can determine the decay time of
the Skyrmion when it is exactly overlapped with the monopole. In two
spatial dimensions, one can study Skyrmion-monopole collisions at zero
impact parameter; however, the catalysis cross-section can be
established only by studying full-3D geometry. This is a major
challenge, taking into account that the Skyrme dynamics is known
\cite{Allder:1987kq}--\cite{Battye:1996nt} to become unstable already in 2D simulations, and a number of
similar problems such as full numerical study of Skyrmion-anti-Skyrmion annihilation remain so
far unsolved.

\section{Acknowledgements}
This work is done in collaboration with Y.~Brihaye, V~A.~Rubakov and D.~H.~Tchrakian.

\end{document}